# Simultaneous whole-animal 3D-imaging of neuronal activity using light-field microscopy

Robert Prevedel[1-3,10], Young-Gyu Yoon[4,5,10], Maximilian Hoffmann,[1-3], Nikita Pak[5,6], Gordon Wetzstein[5], Saul Kato[1], Tina Schrödel[1], Ramesh Raskar[5], Manuel Zimmer[1], Edward S. Boyden[5,7-9] and Alipasha Vaziri[1-3]

[1]Research Institute of Molecular Pathology, Vienna, Austria.
[2]Max F. Perutz Laboratories, University of Vienna, Vienna, Austria.
[3]Research Platform Quantum Phenomena & Nanoscale Biological Systems (QuNaBioS), University of Vienna, Vienna, Austria.
[4]Department of Electrical Engineering and Computer Science, Massachusetts Institute of Technology (MIT), Cambridge, MA, USA
[5]MIT Media Lab, Massachusetts Institute of Technology (MIT), Cambridge, MA, USA
[6]Department of Mechanical Engineering, Massachusetts Institute of Technology (MIT), Cambridge, MA, USA
[7]Department of Biological Engineering, Massachusetts Institute of Technology (MIT), Cambridge, MA, USA
[8]Department of Brain and Cognitive Sciences, Massachusetts Institute of Technology (MIT), Cambridge, MA, USA
[9]McGovern Institute, Massachusetts Institute of Technology (MIT), Cambridge, MA, USA

[10] These authors contributed equally to this work.

Correspondence should be addressed to E.S.B. (esb@media.mit.edu) or A.V. (vaziri@imp.ac.at).

**High-speed large-scale 3D imaging of neuronal activity poses a major challenge in neuroscience. Here, we demonstrate simultaneous functional imaging of neuronal activity at single neuron resolution in an entire *Caenorhabditis elegans* and in larval zebrafish brain. Our technique captures dynamics of spiking neurons in volumes of ~700 $\mu$m x 700 $\mu$m x 200 $\mu$m at 20 Hz. Its simplicity makes it an attractive tool for high-speed volumetric calcium imaging.**

To understand how sensory inputs are dynamically mapped onto functional activity of neuronal populations and how their processing leads to cognitive functions and behavior, tools for non-invasive interrogation of neuronal circuits with high spatio-temporal resolution are required [1, 2]. A number of approaches for 3D neural activity imaging, taking advantage of chemical and genetically encoded fluorescent reporters [3, 4] exist. While some are based on scanning the excitation light in a volume, either sequentially [5-7] or randomly [8, 9], others try to capture 3D image data simultaneously by mapping axial information onto a single lateral plane using a range of approaches [10-14].

Light-field microscopy (LFM) [12] is one such simultaneous 3D imaging method which has been applied to still and *in-vitro* biological samples [12, 13]. In contrast to conventional imaging schemes a light-field microscope captures both the 2D location and 2D angle of the incident light. This is done by placing a microlens array in the native image plane such that sensor pixels capture the rays of the light-field simultaneously. Such 4D light-fields allow the synthesis of a focal stack computationally. Since in LFM single sensor images are used to retrieve the entire 3D volume information, this enables high-speed volumetric acquisition. However, despite its potentially superb temporal resolution, LFM has not to date been used for functional biological imaging. This is because capturing the 4D light-field information by a single sensor image comes at the cost of reduced spatial resolution, and because of inherent trade-offs between axial imaging range and the spatial and axial resolution[12].

Here, we report that neural tissues expressing calcium sensors can be imaged at volume rates of up to 50 Hz and at single neuron resolution, using a 3D deconvolution algorithm [15, 16] applied to LFM. We achieve effective resolutions up to $\sim$1.4 $\mu$m and 2.6 $\mu$m in the lateral and axial dimensions respectively, inside biological samples. To realize our light-field deconvolution microscope (LFDM), we placed a microlens array at the image plane of an epi-fluorescence microscope (**Fig. 1a** and Methods), which captured the different perspectives of the sample (**Fig. 1b**) on the camera sensor. To overcome the trade-off between axial and lateral spatial resolution in LFM [12] we exploited aliasing of the recorded data and used computational reconstruction methods based on 3D-deconvolution to effectively obtain improved lateral and axial resolution [15, 16] (see Methods and **Supplementary Note 1 & 2** for details).

To evaluate the spatial resolution of our LFDM we imaged sub-diffraction beads and reconstructed the point-spread function (PSF) of our system (**Fig. 1b-c**). Using a 40x objective we found a resolution of $\sim$1.4 $\mu$m (2.6 $\mu$m) in the lateral (axial) dimension. To verify the suitability of LFDM for capturing the activity of individual neurons, we imaged a sample consisting of 6 $\mu$m-sized fluorescent beads randomly distributed in three dimensions in agarose and compared a conventional focal stack (taken without microlenses) (**Fig. 1d-e**) with the deconvolved light-field images (**Fig. 1f-g**).

Using the same objective we were able to image the majority of *C. elegans* ($\sim$350 $\mu$m x 350 $\mu$m x 30 $\mu$m) while maintaining single-neuron resolution (**Fig. 2a-c**; **Supplementary Fig. 1-4** and **Supplementary Video 1-5**). We could record activity of neurons in the brain region surrounding the nerve ring and the ventral cord at 5 Hz volume rate. We note that our LFDM allows for substantially higher volume rates which we demonstrated by recording unrestrained worms at 50 Hz (**Supplementary Fig. 4** and **Supplementary Video 3**). Such volume rates would in principle be sufficient for performing whole brain imaging in freely moving worms, especially if additional tracking is employed as previously shown for single neurons [17]. However, since $Ca^{2+}$-signals in *C. elegans* occur typically at timescales of up to 1 Hz, we chose slower volume rates (5 Hz) in order to maximize signal-to-noise and reduce potential photo-bleaching.

The wide FOV of LFDM and the intrinsic simultaneity of the acquisition allow one to study correlations in activity of neurons that are at some distance from each other, which would not be feasible with other unbiased $Ca^{2+}$-imaging techniques. In our experiments, we observed correlated and anti-correlated activity patterns between the pre-motor interneurons in the head and motor neurons located along the ventral nerve cord, which connect to body wall muscles according to the wormatlas (**Fig. 2a-c**).

We used the location, morphology and activity patterns of some of these neurons to identify specific pre-motor interneuron classes such as AVA, AVE, RIM, AIB, AVB, and A- and B-class motor neurons, that have been associated with motor program selection (**Supplementary Fig. 3**) [18]. AVA neurons have been associated with a switch from forward to backward directed crawling, which depends on A-class motor neurons [19], and is associated with a change in the relative activities of A- and B-class motorneurons [18]. Consistent with these findings, we observed high correlation of AVA and A-class motor neuron activity and an anti-correlation of AVA and B-class motor neuron activity. Further, we demonstrated that sensory stimulation can be used to identify neuron classes (**Supplementary Fig. 3**). Applying consecutive 30 second shifts between high and low oxygen levels we observed two neuron classes with increasing $Ca^{2+}$-transients upon oxygen up- and downshift, respectively. Morphology, location and activity patterns of these neuron classes match with those of the oxygen chemosensory neurons BAG and URX [5].

We also recorded exclusively from brain regions surrounding the nerve ring (**Fig. 2d-f,** and **Supplementary Fig. 2**). Imaging smaller FOVs ($\sim$200 $\mu$m x 70 $\mu$m x 30 $\mu$m) leads to faster volume reconstructions and less artifacts stemming from brightly fluorescing cells, such as coelomocytes. Similar to previous findings [5], we were able to resolve up to 74 individual neurons in a typical recording, of which around 30 showed pronounced activity over the recording time of 200 seconds (**Fig. 2d-f**, and **Supplementary Fig. 2**).

In order to highlight the temporal resolution and the broader applicability of our technique for capturing dynamics of large populations of spiking neurons we performed $Ca^{2+}$-imaging in live zebrafish larvae brains expressing GCaMP5 pan-neuronally. Employing a 20x objective, we demonstrated whole-brain $Ca^{2+}$ imaging for volumes spanning $\sim$700 $\mu$m x 700 $\mu$m x 200 $\mu$m at a 20 Hz volume rate. Although in this case optical single-cell resolution had to be compromised in favor of larger FOVs, we could still recover spatially resolved cellular signals over the entire time series using standard signal extraction and unmixing techniques [20]. Implementing this approach we extracted neuronal activity for $\sim$5000 cells across the brain, and followed their fast $Ca^{2+}$-transients on millisecond timescale (**Fig. 3** and **Supplementary Video 6**). By applying an aversive odor to the fish we evoked neuronal activity and inferred dynamics of $Ca^{2+}$-signals across the olfactory system, the midbrain and parts of the hindbrain consistent with previous demonstrations of the neuronal dynamics in these regions [6, 7, 21-23]. The high time-resolution revealed subtle differences in the exact timing of the onset of the response for different cells. Groups of cells constituting of neurons all within similar spatial proximity can be observed (**Fig. 3c**). While the neurons in each group exhibit a nearly synchronous onset of their activity, the collective response of each group is delayed with respect to the others. Our imaging speed, more than an order of magnitude faster than in previous whole-brain functional imaging [6, 7], is thus able to reliably capture the dynamic activity of a large number of cells with high spatial and temporal resolution.

In summary, we have implemented a LFDM and demonstrated its ability to simultaneously capture the neuronal activity of the entire nervous system of *C. elegans* at single cell resolution as well as record dynamics of spiking neurons by performing whole-brain $Ca^{+2}$ imaging in larval zebrafish at 20 Hz. The increase in spatial resolution compared to LFM was achieved by performing deconvolution during post-processing. The simultaneity of acquisition of volumes in LFDM eliminates spatio-temporal ambiguity associated with sequentially recorded approaches and decouples temporal resolution from volume size. Resolutions in all three dimensions are set by the objective and microlens properties, while FOV and acquisition rate are determined by the camera chip size, frame rates and signal intensity. LFDM is easy to set-up, cost-effective and compatible with standard microscopes. Both the temporal resolution and the obtainable FOVs make LFDM an attractive technique for future combination with behavioral studies. Future work will focus on obtaining higher spatial resolutions and larger FOVs as well as faster and more efficient computational reconstruction techniques, both of which of are expected to improve with technological advancements in camera sensors and processors. Finally, the use of red-shifted $Ca^{2+}$-sensors [24] and the combination of LFDM with techniques for imaging at depth in biological tissue [25] bears further potential for wide-spread use of this method.

## ACKNOWLEDGMENTS

We thank T. Müller, P. Pasierbek, P. Forai, H. Kaplan, M. Molodtsov, K. Tessmar-Raible and F. Schlumm for technical support and loan of equipment, as well as H. Baier (MPL for Neurobiology, Munich) and M. Orger (Champalimaud, Lisbon) for sharing zebrafish lines. The computational results presented have been achieved in part using the Vienna Scientific Cluster (VSC). This work was supported by the VIPS Program of the Austrian Federal Ministry of Science and Research and the City of Vienna as well as the European Commission (Marie Curie, FP7-PEOPLE-2011-IIF) (R.P.), a Samsung Scholarship (Y.-G.Y), an NSF Graduate Fellowship (N.P.), the Allen Institute for Brain Science, the MIT Media Lab, the MIT McGovern Institute, NIH 1R01EY023173, the MIT Synthetic Intelligence Project, the IET Harvey Prize, NSF CBET 1053233, the New York Stem Cell Foundation-Robertson Award, NSF CBET 1344219, NIH 1DP1NS087724, Google, the NSF Center for Brains Minds and Machines at MIT, and Jeremy and Joyce Wertheimer (E.S.B.), the Vienna Science and Technology Fund (WWTF) project VRG10-11, Human Frontiers Science Program Project RGP0041/2012, Research Platform Quantum Phenomena and Nanoscale Biological Systems (QuNaBioS) (A.V.), the Research Institute of Molecular Pathology (IMP) and the European Community's Seventh Framework Programme / ERC no. 281869)(M.Z. and T.S.). The IMP is funded by Boehringer Ingelheim.

## AUTHOR CONTRIBUTIONS

## COMPETING FINANCIAL INTERESTS

The authors declare no competing financial interests.

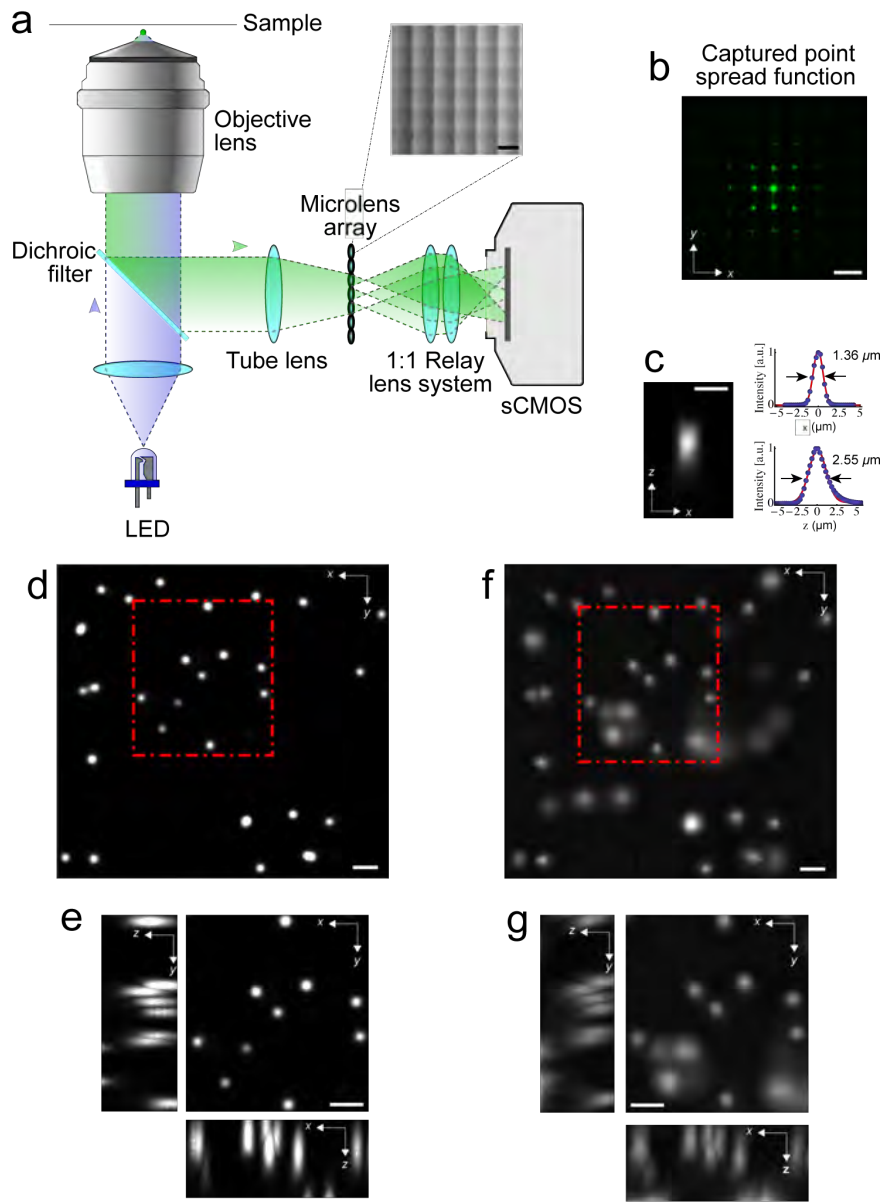

**Figure 1 |** Light-field deconvolution microscopy (LFDM). (**a**) A microlens array was appended to the camera port of a wide-field microscope and placed in the primary image plane of the fluorescence microscope. The array itself was imaged with a 1:1 relay lens system onto the chip of a sCMOS camera. (Methods) The inset shows a close-up picture of the microlens array. (**b**) The point spread function (PSF) of a sub-diffraction sized bead located at $z = 7.5$ $\mu$m off the focal plane, as seen through the microlens array. Scale bar 150 $\mu$m in (**a, b**). (**c**) Axial (*xz*) PSF at $z = 7.5$ $\mu$m, reconstructed using LFDM, and corresponding *x*- and *z*-profiles, showing lateral and axial resolution, respectively. Scale bar 3 $\mu$m. (**d**) Maximum-intensity projection of a deconvolved wide-field focal stack taken without microlenses. The sample consists of 6 $\mu$m-sized fluorescent beads in agarose. (**e**) Zoom-in of highlighted area; *xz*- and *yz*-section maximum-intensity projections are shown. (**e, f**) Corresponding volume of the same beads, 4 – 28 $\mu$m off the focal plane, reconstructed via 15 iterations of light-field deconvolution algorithm. Scale bar 10 $\mu$m in (**d-g**).

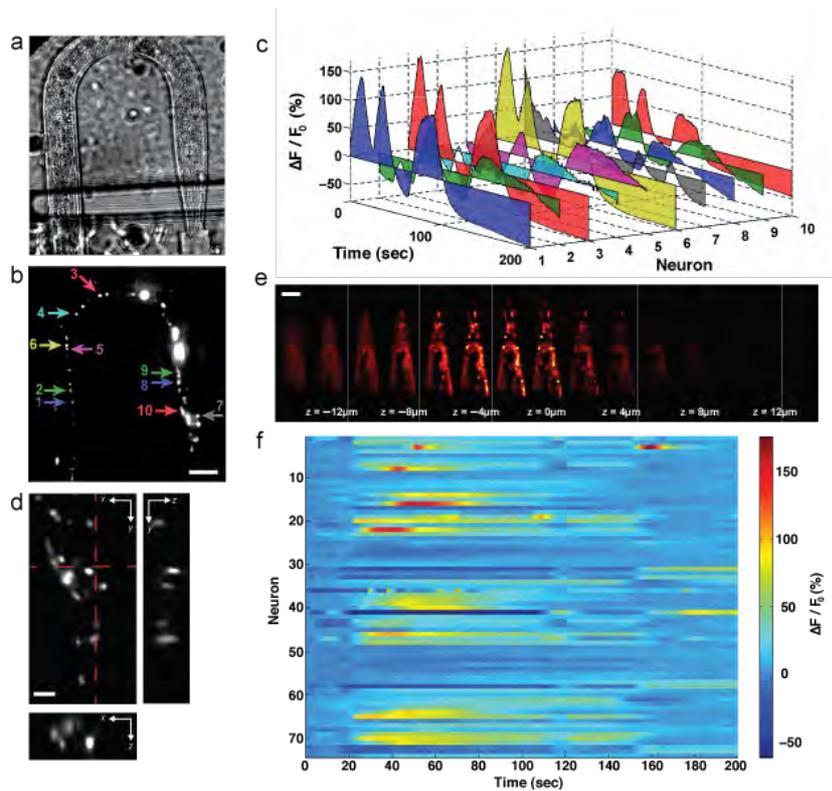

**Figure 2 |** Whole animal Ca²⁺-imaging of *Caenorhabditis elegans* using LFDM. (**a**) Wide-field image of the worm inside a microfluidic poly(dimethylsiloxane) (PDMS) device used for immobilization. Head is at bottom right. (**b**) Maximum intensity projection (MIP) of light-field deconvolved image (15 iterations) containing 14 distinct *z*-planes. Arrows indicate individual neurons in head ganglia and ventral cord. Scale bar 50 $\mu$m. (**c**) Ca²⁺-intensity traces ($\Delta$F/F$_0$) of NLS-GCaMP5K fluorescence of selected neurons as marked in (**b**), and imaged volumetrically at 5 Hz for 200 seconds. Also see **Supplementary Video 1**. (**d**) Zoom of the brain region, with MIP of *xy* plane as well as *xz* and *yz* cross-sections indicated by the dashed lines. (**Supplementary Video 2**). Scale bar 10 $\mu$m. (**e**) Individual z planes of typical recording of the worm's brain, reconstructed from a single camera exposure (see **Supplementary Fig.** 2 for neuron IDs). In this recording, the worm's center along the lateral (left-right) axis was placed at the focal plane of the objective. Scale bar 50 $\mu$m. (**f**) Activity of all 74 neurons identified in (**e**), also see **Supplementary Video 4**. Each row shows a time-series heat map of an individual neuron. Color indicates percent fluorescence changes ($\Delta$F/F$_0$); scaling is indicated by the color bar on the right.

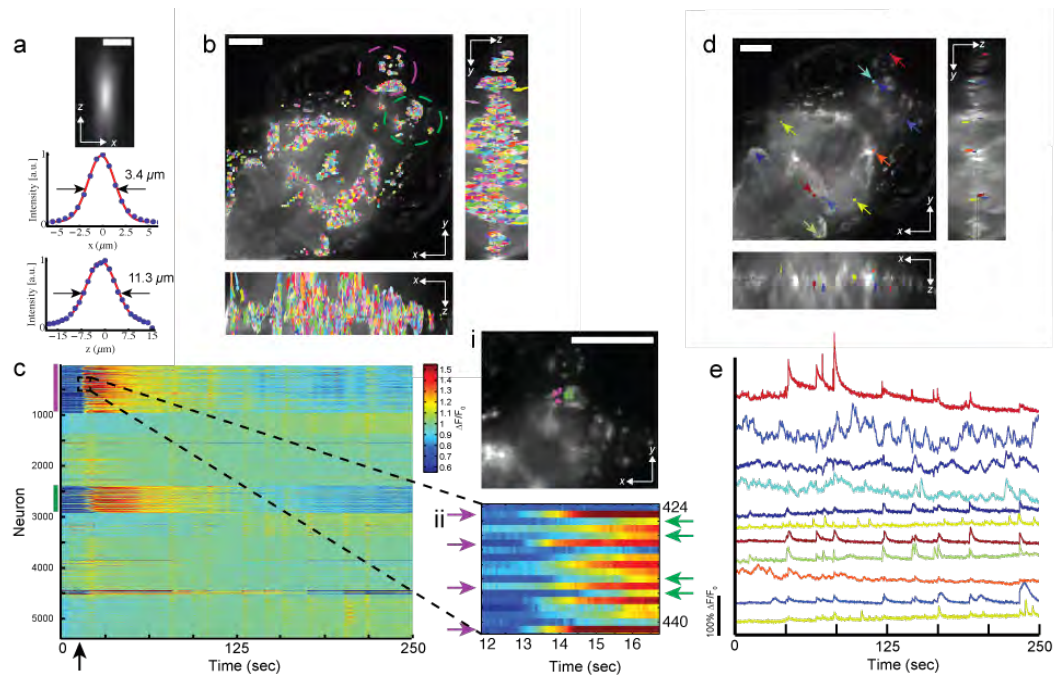

**Figure 3 |** Whole-brain Ca$^{2+}$-imaging of larval zebrafish *in vivo*.
(**a**) Axial point spread function (PSF) of a 0.5 $\mu$m sized bead located at z = 28 $\mu$m off the focal plane for the 20x 0.5NA lens and corresponding *x* and *z*-profiles. Scale bar 10 $\mu$m. (**b**) Maximum intensity projection (MIP) of light-field deconvolved volume (8 iterations) containing 51 *z* planes, captured at 50ms per frame exposure time and spaced 4 $\mu$m apart, showing *xy* plane, *xz* and *yz* cross-sections. Spatial filters, each representing individual cells, identified using PCA/ICA analysis [20] are shown. In total, 5,379 filters were automatically identified most of which correspond to individual neurons. (**c**) Extracted Ca$^{2+}$-intensity signal ($\Delta$F/F$_0$) of GCaMP5 fluorescence using spatial filters shown in (**b**). Each row shows a time-series heat map. Color bars denote encircled regions in (**b**), which include olfactory epithelium, olfactory bulb and telencephalon. The arrow at ~15 sec denotes the addition of an aversive odor. A zoom-in of the dashed area is shown (right, lower panel), neurons with subtle differences in response onset are highlighted by colored arrows. Their location in the MIP is also shown (right, upper panel). (**d**) Overlay of MIP with randomly selected spatial filters. (**e**) Ca$^{2+}$-intensity traces of selected cells shown in (**d**). Neurons were manually selected from the olfactory system, midbrain and hindbrain. Trace color matches spatial filter color in (**d**). Scale bar is 100 $\mu$m in (**b**), (**d**) and (**e**). Also see **Supplementary Video 6.**

## Online Methods

### Setup

The LFM system is appended to an epi-fluorescence microscope (Zeiss, Axiovert 200) equipped with a LED excitation light source ($\lambda$ = 488nm, 300 mW, CoolLED) and a standard GFP filter set (Zeiss). In all *C. elegans* imaging experiments, we used a 40x 0.95NA dry objective (Zeiss APOCHROMAT), while zebrafish imaging was performed with a 20x 0.5NA dry objective (Zeiss Plan-NEOFLUAR). The microlens array is mounted inside a 5-axis kinematic mount (Thorlabs) to allow fine adjustment of the array orthogonal to the optics axis, which we found crucial for high-quality results. The array is further imaged onto a 5.5 Megapixel (2560 x 2160 pixels) sCMOS camera (Andor Zyla) utilizing a 1:1 relay macro lens objective (Nikon AF-S 105mm 2.8 G VR IF-ED Micro) (**Fig. 1a**). Details on optical design choices and their effect on resolution are discussed in **Supplementary Note 1**.

### *C. elegans* experiments

To record neuronal activity from *C. elegans*, we loaded adult worms (one- to four-egg-stage) expressing nls-GCaMP5K under the *punc31*-promoter (strains ZIM294 and ZIM617) into a microfluidic channel which was connected to a reservoir containing S-Basal buffer with 5 mM tetramisole, an acetylcholine receptor–specific agonist that mildly paralyzes the animal's muscles to reduce motion [5]. The worm was placed off the native focal plane and towards the objective using a piezo stepper motor (PI-721, Physik Instrumente GmbH), such that the entire worm was ideally contained in the region spanning -30 $\mu$m to 0 $\mu$m. By doing so we exploited the highest resolution of LFDM while avoiding artifacts near the focal plane. When recording from the head region only, the worm's head ganglia were placed at the center of the FOV and excitation was limited to this area by the use of an iris in the excitation pathway. For the experiments involving chemosensory stimulation, we followed the procedure described in ref. [5]. Neurons were identified by classification according to size, shape and relative positions of cell nuclei [28] previously described characteristic activity patterns [5] were used as further confirmations. AVA neurons are located in the anterior-ventral part of the lateral ganglia and exhibit an elongated nucleus. AVE neurons are situated posteriorly-medially to AVA and have a similar activity pattern [18]. RIM neurons are located in the posterior ventral part of the lateral ganglia; their position is often ambiguous with RIB and AIB neurons, which also exhibit activity patterns similar to RIM. VB01 is located in the anterior to middle part of the retrovesicular ganglion, its position is ambiguous with other motorneurons in this region, like DB02. DA01 is located at the posterior end of the retrovesicular ganglion. AVB neurons are located central to the lateral ganglia and typically show anticorrelated activity with AVA. Ambiguities are posed by the nearby neurons AIN, AVD, AVH and AVJ. BAG neurons are located at the posterior end of the anterior ganglion and exhibit the largest cell nucleus in this region; they reliably respond to oxygen downshift. URX neurons are located at the anterior dorsal end of the lateral ganglia directly ventrally to the unambiguously identifiable nucleus of ALA. URX neurons reliably respond to oxygen upshift.

**Zebrafish larvae experiments**

For zebrafish experiments, *mitfa⁻/⁻* larvae with pan-neuronal GCaMP5 expression have been imaged 5-8 dpf (days post fertilization) using stable lines *HuC:GCaMP5G* and *HuC:Gal4/UAS:GCaMP5G*, respectively. The fish were immobilized by embedding them in 2% agarose with their mouth and tail cleared of agarose to allow for odor stimulation and tail movement. Odor stimulation was performed by manually supplying water containing highly concentrated rotten fish extract into the imaging chamber during recordings.

**Light-field deconvolution**

The volume reconstruction itself can be formulated as a tomographic inverse problem [26], where multiple different perspectives of an 3D volume are observed, and linear reconstruction methods—implemented via deconvolution—are employed for computational 3D volume reconstruction. The image formation in light-field microscopes involves diffraction from both the objective and microlenses. Point spread functions for the deconvolution can be computed from scalar diffraction theory [27]. More details are given in **Supplementary Note 2**.

After recording the raw light-field images, the digital images were cropped to regions of interest and resampled to contain 11 x 11 or 15 x 15 angular light-field samples under each lenslet. Two calibration images, one showing the microlenses with collimated rear-illumination and one showing a uniform fluorescent slide, were used for digital image rectification, in which camera pixels are assigned to individual microlenses. Reconstruction of each frame of an image sequence took between 2 and 30 minutes, depending on the size of the image, number of iterations of the deconvolution algorithm, reconstruction method and workstation used. Computational resources are further discussed in **Supplementary Note 2**.

**Ca²⁺ imaging data analysis**

To extract a fluorescence time series for individual neurons from the 4D data, different strategies were employed for *C. elegans* and zebrafish. For *C. elegans*, we first applied rigid-body motion correction to each individual z plane movie. We then computed a median-intensity projection through time for each motion-corrected *z* plane movie and used a maxima-finding algorithm to identify areas of peaked intensity in each projection. A circular region of interest (ROI) was created surrounding each intensity peak, and overlapping ROI areas within *z* planes were eliminated. ROIs in adjacent *z* planes within an *x-y* distance of 7 pixels were considered to be a component of the same neuron, up to a maximum of 5 planes, and for each neuron at each time point, the brightest 100 pixels of the aggregate of all pixels within the neuron's component ROIs were averaged to produce a single fluorescence value and de-trended with an exponential decay function to account for photobleaching. For zebrafish, the data was first de-trended based on the overall intensity of each frame. Then, to reduce time series data, first in-active voxels were discarded based on their time-domain variance. Splitting the volume into smaller sub-volumes further reduced data size. We followed the strategy proposed in reference [20] to extract cellular signals from the Ca²⁺-imaging data. Each sub-volume data underwent PCA/ICA for automated spatial filter extraction where ideally each spatial filter corresponds to the location of a neuron [20]. After automatically rejecting spatial filters based on size and dispersion, we applied the spatial filters to the 4D data to extract their fluorescence intensity. Time-points during which the fish seemed to contract where discarded and replaced with nearest neighbor fluorescence intensities. These contractions typically lasted between 200 ms and 1 sec only and were temporally very sparse. Therefore we regarded them negligible compared to the overall recording time. Fish which moved substantially during image acquisition were discarded from analysis. To extract $\Delta F/F_0$, we calculated $\Delta F/F_0 = 100 * (F_{(t)} - F_0)/F_0$, with $F_0$ being the mean fluorescence intensity of each corrected trace.

**Methods-only references**

# Supplementary Information

# Simultaneous whole-animal 3D-imaging of neuronal activity using light field microscopy

Robert Prevedel[1-3,10], Young-Gyu Yoon[4,5,10], Maximilian Hoffmann,[1-3], Nikita Pak[5,6], Gordon Wetzstein[5], Saul Kato[1], Tina Schrödel[1], Ramesh Raskar[5], Manuel Zimmer[1], Edward S. Boyden[5,7-9] and Alipasha Vaziri[1-3]

[1]Research Institute of Molecular Pathology, Vienna, Austria.
[2]Max F. Perutz Laboratories, University of Vienna, Vienna, Austria.
[3]Research Platform Quantum Phenomena & Nanoscale Biological Systems (QuNaBioS), University of Vienna, Vienna, Austria.
[4]Department of Electrical Engineering and Computer Science, Massachusetts Institute of Technology (MIT), Cambridge, MA, USA
[5]MIT Media Lab, Massachusetts Institute of Technology (MIT), Cambridge, MA, USA
[6]Department of Mechanical Engineering, Massachusetts Institute of Technology (MIT), Cambridge, MA, USA
[7]Department of Biological Engineering, Massachusetts Institute of Technology (MIT), Cambridge, MA, USA
[8] Department of Brain and Cognitive Sciences, Massachusetts Institute of Technology (MIT), Cambridge, MA, USA
[9] McGovern Institute, Massachusetts Institute of Technology (MIT), Cambridge, MA, USA

[10] These authors contributed equally to this work.

Correspondence should be addressed to E.S.B. (esb@media.mit.edu) or A.V. (vaziri@imp.ac.at).

**Supplementary Figures**

**Supplementary Figure 1.** Whole-animal Ca²⁺-imaging of *C. elegans*.

**Supplementary Figure 2.** High-resolution images of Fig. 2e and Fig. 2f indicating Neuron ID numbers in z-planes and heatplot map of neuronal activity of all neurons.

**Supplementary Figure 3.** Identification of neuron classes in *C. elegans* during chemosensory stimulation.

**Supplementary Figure 4.** High-speed Ca²⁺-imaging of unrestrained *C. elegans*.

**Supplementary Note 1** General principle, optical design choices and their effect on resolution in 3D deconvolution light field microscopy.

**Supplementary Note 2** Volume reconstruction for 3D-deconvolution light field microscopy and computing requirements.

**Supplementary References**

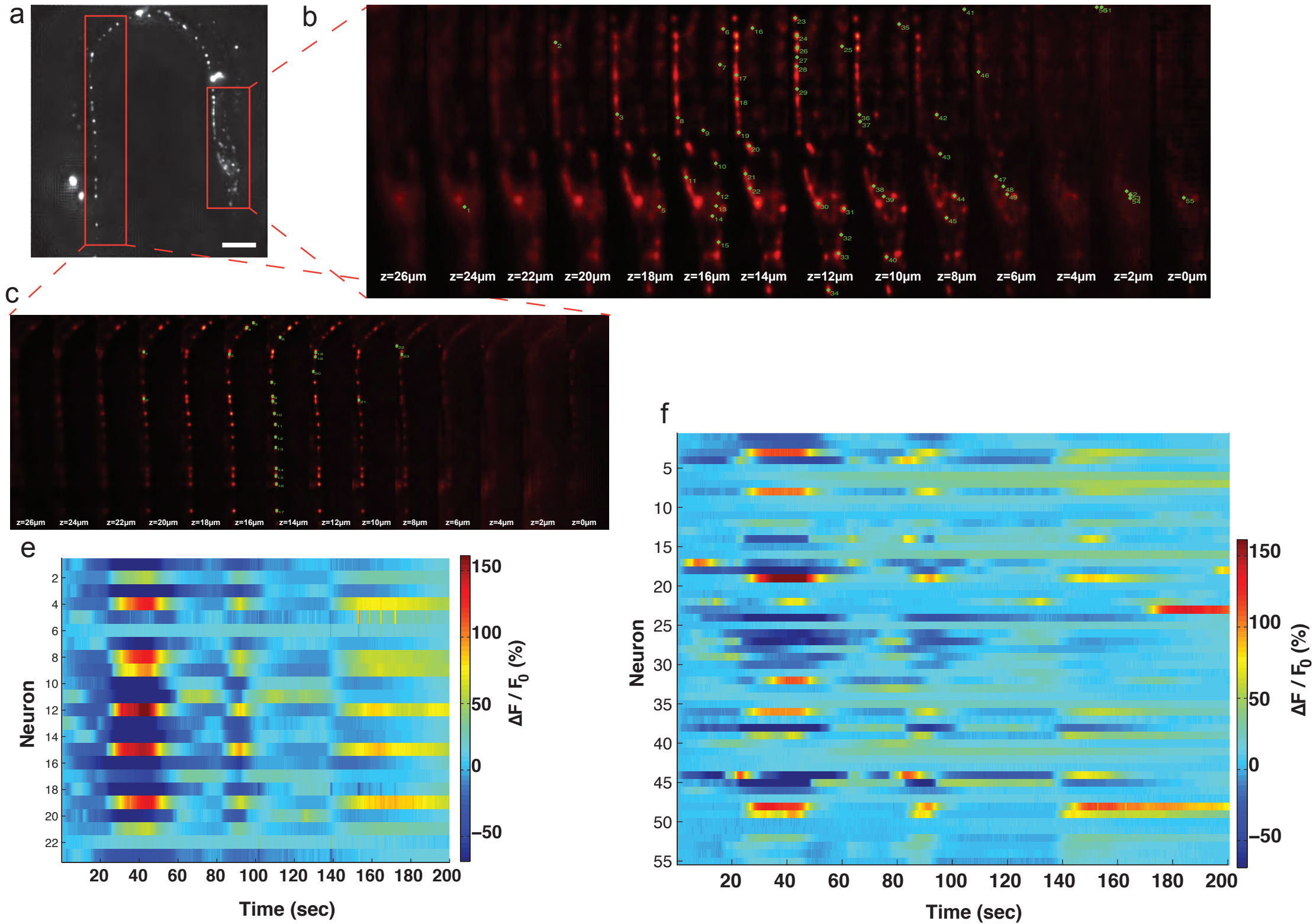

**Supplementary Figure 1.** Whole-animal $Ca^{2+}$-imaging of *C. elegans*.

(**a**) Maximum intensity projection (MIP) of light field deconvolved image (15 iterations) of the whole worm shown in Fig. 2d, containing 14 distinct $z$- planes. Neurons contained in red boxes were further analyzed in (**b-f**). NeuronIDs of $z$-stack in **b** match with heatmap plot of neuronal activity in **f** and show neurons identified in the head using an automated segmentation algorithm, while **c** shows neuronIDs along the ventral cord with corresponding heatplot map shown in **e**. Scale bar 50 $\mu$m.

a

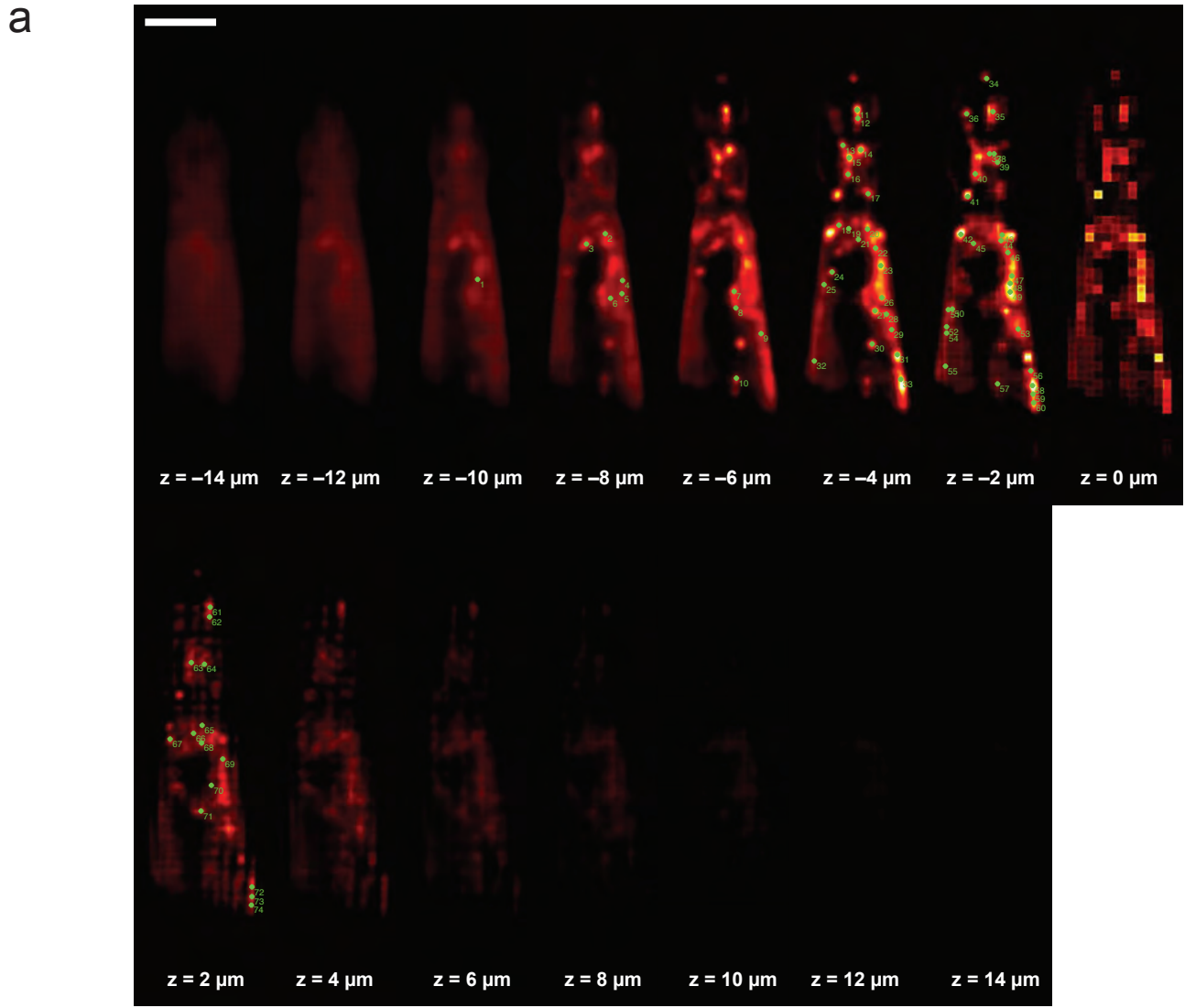

b

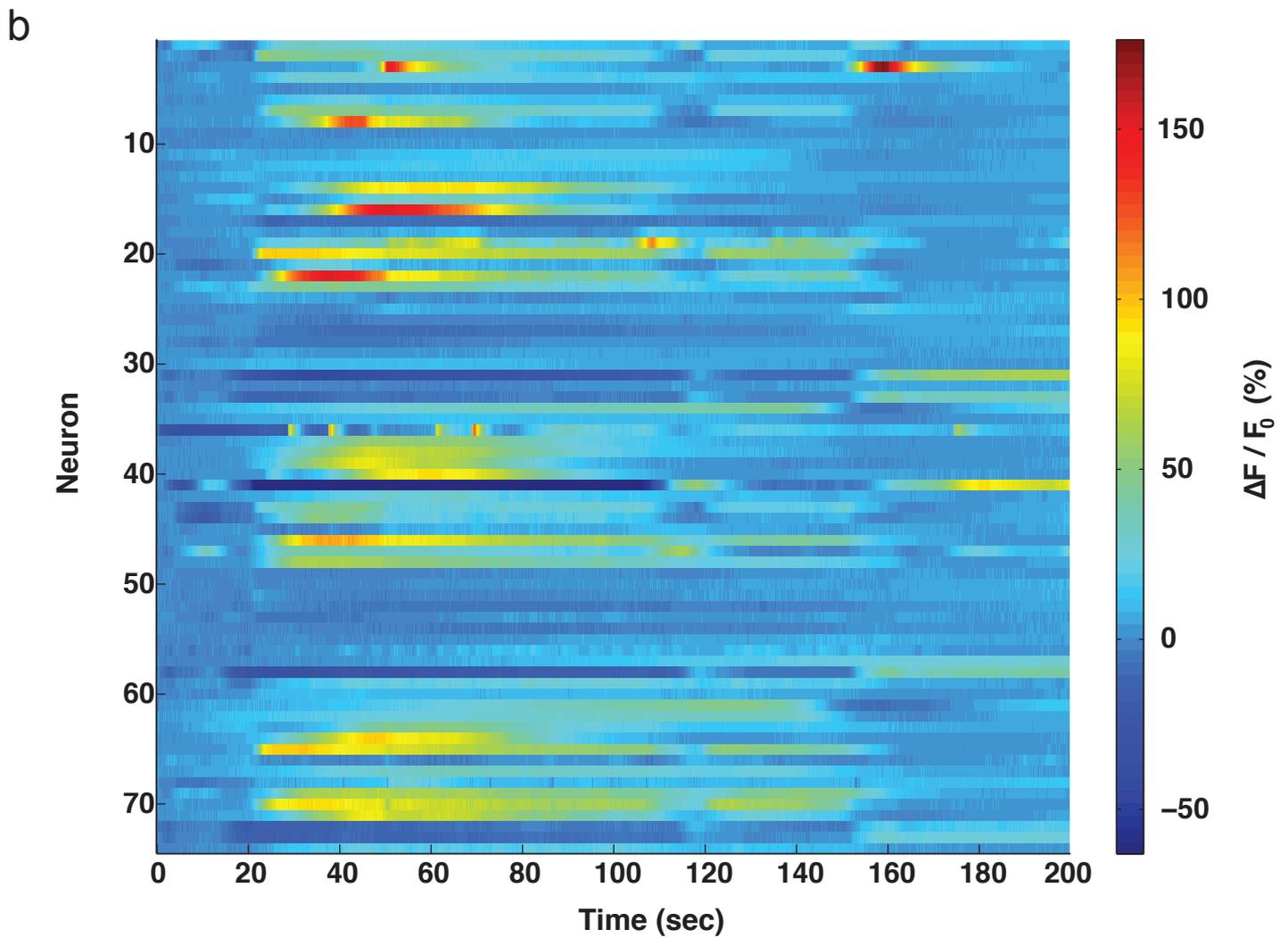

**Supplementary Figure 2.** High-resolution images of Fig. 2e and Fig. 2f indicating Neuron ID numbers in *z*-planes in (**a**) and heatmap plot of neuronal activity of all neurons in (**b**).

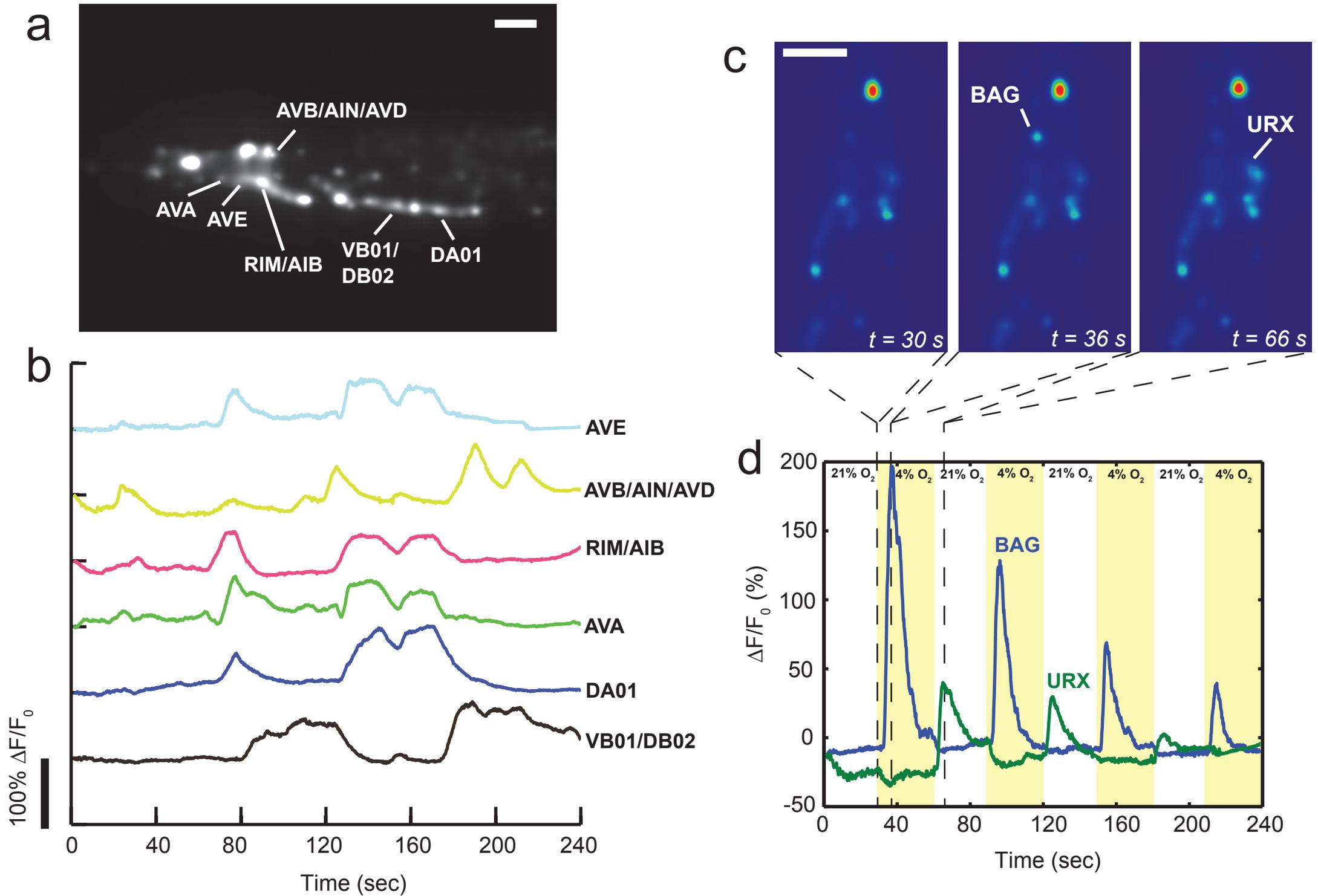

**Supplementary Figure 3.** Identification of neuron classes in *C. elegans* during chemosensory stimulation.

**Supplementary Figure 3.** Identification of neuron classes in *C. elegans* during chemosensory stimulation.

Whole brain LFDM recording at 5 Hz of *C. elegans* under consecutively changing $O_2$ concentrations (30 seconds time-shifts). (**a**) Maximum intensity projection (MIP) of light field deconvolved image (8 iterations) of the worm's head region, containing 7 distinct *z*-planes. Neuron classes were identified based on location and typical $Ca^{2+}$-signals, whose individual traces are shown in **b**. (**c**) Individual *z*-plane containing the oxygen-downshift sensing neuron BAG at various time-points before, during and after stimulus, respectively. (**d**) Fluorescence traces of oxygen sensory neurons BAG and URX, with varying $O_2$ concentrations indicated by shading. Scale bar is 20 $\mu$m in **a** and **c**.

**Supplementary Figure 4.** High-speed Ca²⁺-imaging of unrestrained *C. elegans* at 50 Hz.

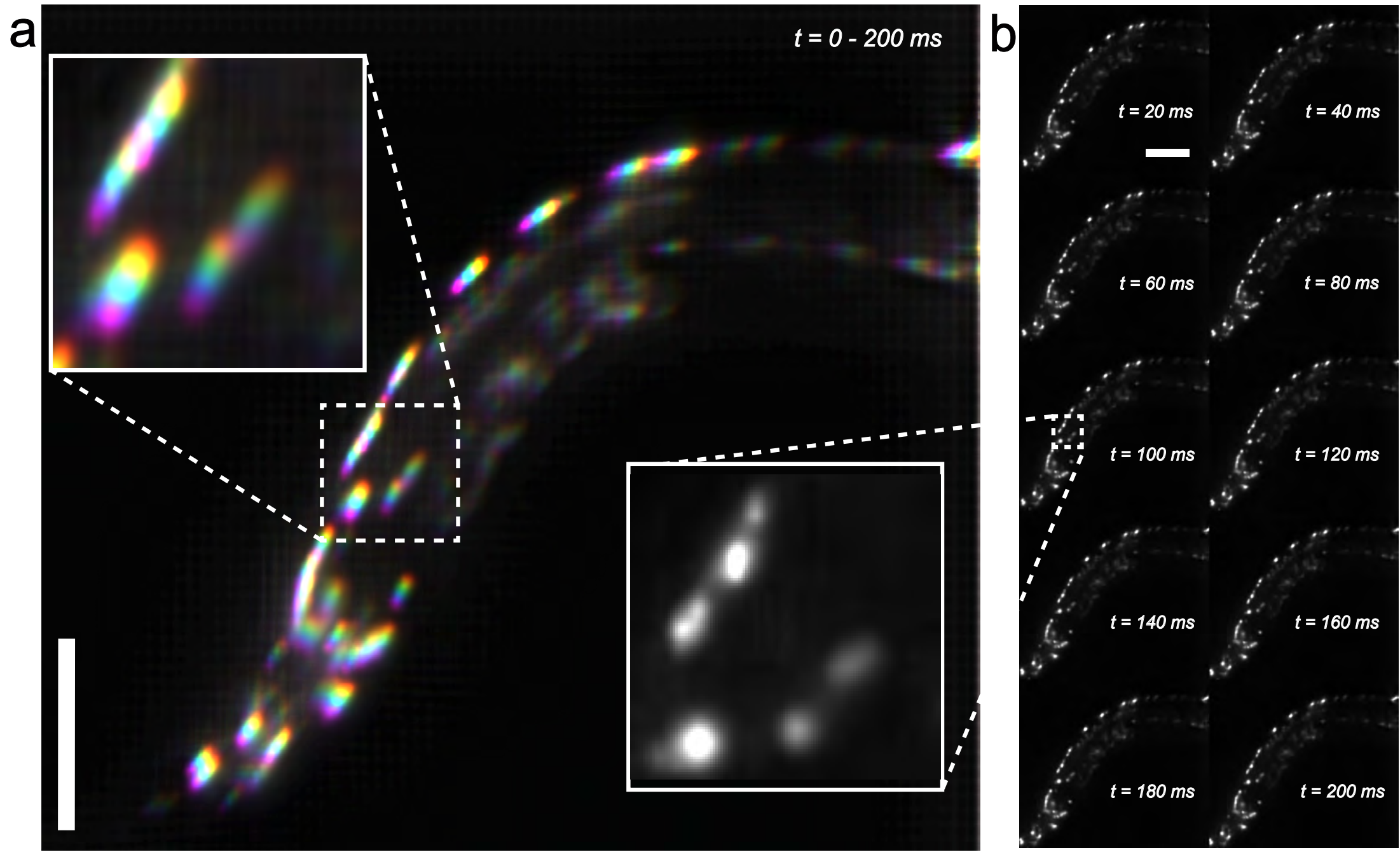

**Supplementary Figure 4.** High-speed $Ca^{2+}$-imaging of unrestrained *C. elegans* at 50 Hz. Selected time-series of the LFDM recording of freely-moving worms at 50 Hz shown in **Supplementary Video 4**. (**a**) Overlay of 10 consecutive frames, with colors coding for different time-points. This is equivalent to an effective frame-rate of 5 Hz. At this speed, motion blur would lead to ambiguous discrimination of individual neurons, as is clearly visible in the inset. In contrast, in (**b**) we show the individual frames of the same time-series as recorded with 50 Hz (20 ms exposure time). At this speed, motion blur is almost non-existent. This demonstrates that 50 Hz are sufficient to follow the activity of unrestrained worms, especially if additional worm tracking would be employed. Scale bar is 50 $\mu$m in **a** and **b**. Also see **Supplementary Video 3**.

**Supplementary Note 1** General principle, optical design choices and their effect on resolution in 3D deconvolution light field microscopy.

Generally speaking, a conventional 2-D microscope captures a high-resolution image of a specimen that is in focus. For volumetric specimens, the same image, however, also contains blurred contributions of areas that are optically out of focus. Unmixing them in post-processing is an ill-posed problem and usually not possible. Scanning microscopes solve this problem by measuring each point in the 3-D volume sequentially. While this is an effective process, it is time-consuming and not always applicable to capturing dynamic events or moving specimens. Light field microscopes change the optical acquisition setup to capture different primitives: instead of recording individual points sequentially, light field microscopes capture "rays" of light, that their summed emission through the 3-D volume. Instead of recording them in sequence, a set of "rays" – the light field – is multiplexed into a single 2-D sensor image. This spatial, rather than temporal, approach to multiplexing drastically improves acquisition speed at the cost of reduced resolution. To recover the 3-D volume from measured emission, a computed tomography problem has to be solved. Following Ref. 1, we implement this reconstruction step as a deconvolution. Please note that while the light field is conceptually comprised of geometric rays, in practice the image formation and inversion also considers diffraction, as discussed in the primary text.

Light field microscopes support all objective magnifications, but usually benefit from a high numerical aperture (NA) and microlenses that are matched with the NA of the employed objective. The choice of objective and microlens array determines the spatial resolution and field-of-view in all three dimensions. The pitch, i.e. the distance between the microlenses, in combination with the sensor's pixel size and objective magnification controls trade-off between spatial resolution vs. field-of-view while the objective's magnification and numerical aperture control axial resolution vs. axial range. Furthermore, the field-number of the microlenses needs to match that of the objective in order to preserve the maximum angular information in the light fields [2].

Due to the variation in sampling density, reconstructed volumes have a lateral resolution that varies along the optical axis. On the focal plane, achievable resolution is equivalent to conventional LFM, i.e. the size of each microlens divided by the magnification of the objective lens (150 $\mu$m / 40x = 3.75 $\mu$m in our system). The resolution increases for lateral sections close to the focal plane, $\sim$1.5$\mu$m laterally in our implementation, but drops at larger distances, e.g. to $\sim$3 $\mu$m laterally at -25 $\mu$m, in accordance with Ref. [1]. We find similar behavior with the 20x 0.5NA lens used in our zebrafish recordings. Here we find a maximum resolution of $\sim$3.4 $\mu$m ($\sim$11 $\mu$m) laterally (axially) based on a reconstructed point spread function (see also **Fig. 3a**).

It is also possible and straightforward to design microlens arrays for higher magnification objectives in order to look at smaller samples. Following the criteria outlined in Ref. 2, microlenses can be designed taking into account the trade-offs between lateral and axial resolution. For instance we have performed simulations for a 100x 1.4NA oil objective and a f-number matched microlens of 100 $\mu$m pitch, and found that our LFDM should have a resolution of $\sim$0.27 $\mu$m (1 $\mu$m) laterally (axially). The lateral field of view would be 140 $\mu$m with a sCMOS camera similar to the one used in this work and we would expect a useful axial range of 10-15 $\mu$m.

**Supplementary Note 2** Volume reconstruction for 3D-deconvolution light field microscopy and computing requirements.

The software for 3D reconstruction was written in MATLAB (Mathworks) using its parallel computing toolbox to enable multi-core processing, and allows choosing between CPU- and GPU-based executions of the algorithm. The software consists of three different parts: point spread function (PSF) computation, image rectification / calibration, and 3D volume reconstruction. To generate PSFs, we compute the wavefront imaged through the microlens array for multiple points in the volume using scalar diffraction theory [3]. We also exploit the circular symmetry of PSF for its computation, which results in a boost in computational speed. To faithfully represent the high spatial frequency component of the wavefront, computations are performed with a spatial oversampling factor of 3x compared to the size of the virtual pixels that correspond to the resampled image.

For the image rectification and calibration, the size and location of each microlens with respect to the sensor pixels are estimated using calibration images showing a fluorescent slide and a collimated beam. An open source software named LFDisplay [http://graphics.stanford.edu/software/LFDisplay/], for example, can be used to locate the microlenses with respect to the pixels. Once the size and the location of each microlens is determined, captured images are resampled to contain 15 x 15 (11 x 11) angular light field samples under each microlens. The target axial resolution of reconstructed volumes is 2 (4) $\mu$m, which requires 12-16 (51) z-slices for worm (zebrafish) samples.

The essential operations for volume reconstruction are based on computing large number of 2-dimensional convolutions. Therefore reconstruction speed depends heavily on the implementation of the convolution operation and its speed. Using the convolution theorem, this problem can be accelerated by computing on graphical processor units (GPUs) in the Fourier domain. The underlying fast Fourier Transform (FFT) can be computed in $O(n \log n)$ operations whereas conventional convolution requires $O(n^2)$ operations. Furthermore, the FFT is well suited for GPU computing, and we found this to result in significant (up to 20x) reduction in computing time compared to 12-core CPU based execution. With GPU computing method, reconstructing individual frames of recorded image sequences using Richardson-Lucy deconvolution method took between 2 and 6 min, depending on the size of the image, on a workstation with one Nvidia Tesla K40c GPU and 128GB of RAM. Specifically, the reconstruction of only the head ganglia region of *C. elegans* (Fig. 2c-e) took about 2 minutes where the reconstruction of the whole C. elegans took about 6 minutes with 8 iterations of the deconvolution algorithm. Similar times were measured for zebrafish volume reconstructions.

In comparison, CPU based computing on 12 parallel cores required between 5 and 30 min. However, by parallelizing the reconstruction on a medium sized cluster employing ~40 nodes, we found that a typical 1000 frame movie of whole *C.elegans* (such as in **Supplementary Video 1**) could be reconstructed within ~12 hours. Cloud based computing options, e.g. through Amazon Web Services and other competing online tools, might also provide efficient means for large-scale volume reconstruction.

Reconstruction times of image sequences could be further optimized by using the reconstructed volume of one frame as the initial guess for the next. This removes the need for multiple algorithmic iterations at each frame and is well-justified because the imaging speed was sufficiently faster than both neuronal activity and movement of the worm.

**Supplementary References**

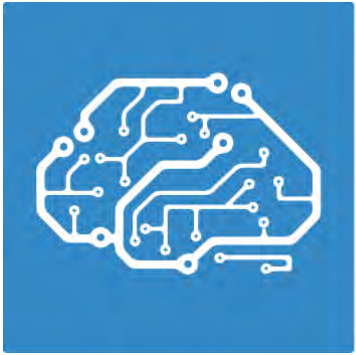

# Center for Brains, Minds & Machines

# Simultaneous whole-animal 3D imaging of neuronal activity using light-field microscopy

by

Robert Prevedel*, Young-Gyu Yoon*, Maximilian Hoffmann, Nikita Pak, Gordon Wetzstein, Saul Kato, Tina Schrödel, Ramesh Raskar, Manuel Zimmer, Edward S Boyden** & Alipasha Vaziri**

**Abstract:** High-speed, large-scale three-dimensional (3D) imaging of neuronal activity poses a major challenge in neuroscience. Here we demonstrate simultaneous functional imaging of neuronal activity at single-neuron resolution in an entire *Caenorhabditis elegans* and in larval zebrafish brain. Our technique captures the dynamics of spiking neurons in volumes of ~700 μm × 700 μm × 200 μm at 20 Hz. Its simplicity makes it an attractive tool for high-speed volumetric calcium imaging. (* equal contributions, ** co-corresponding authors)

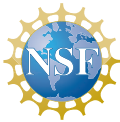 This work was supported by the Center for Brains, Minds and Machines (CBMM), funded by NSF STC award CCF-1231216.